\documentclass[12pt,preprint]{aastex}

\usepackage{emulateapj5}

\newcommand{\BaII}{\ion{Ba}{2}}
\newcommand{\CaII}{\ion{Ca}{2}}

\newcommand{\NII}{\ion{N}{2}}
\newcommand{\OIII}{\ion{O}{3}}
\newcommand{\SrII}{\ion{Sr}{2}}

\newcommand{\kms}{{\>\rm km\>s^{-1}}}

\newcommand{\sprocess}{{\it s}-process}

\shorttitle{Planetary Nebula WeBo 1}
\shortauthors{Bond et al.}

\begin{document}

\title{WeBo 1: A Young Barium Star Surrounded by a Ring-Like Planetary Nebula}


\author{Howard E. Bond\altaffilmark{1},
Don L. Pollacco\altaffilmark{2},
and Ronald F. Webbink\altaffilmark{3}}

\altaffiltext{1}
{Space Telescope Science Institute, 
3700 San Martin Dr.,
Baltimore, MD 21218; 
bond@stsci.edu}

\altaffiltext{2}
{APS Division, Department of Pure and Applied Physics,
Queens University Belfast,
Belfast BT7 1NN,
United Kingdom;
D.Pollacco@Queens-Belfast.ac.uk}

\altaffiltext{3}
{Department of Astronomy,
University of Illinois,
1002 W. Green St.,
Urbana, IL 61801;
webbink@astro.uiuc.edu}

\begin{abstract}

WeBo~1 (PN~G135.6+01.0), a previously unrecognized planetary nebula with a
remarkable thin-ring morphology, was discovered serendipitously on Digitized
Sky Survey images.  The central star is found to be a late-type giant with
overabundances of carbon and \sprocess\ elements. The giant is
chromospherically active and photometrically variable, with a probable period
of 4.7 days; this suggests that the star is spotted, and that 4.7~days is its
rotation period. We propose a scenario in which one component of a binary
system became an AGB star with a dense stellar wind enriched in C and
\sprocess\ elements; a portion of the wind was accreted by the companion,
contaminating its atmosphere and spinning up its rotation. The AGB star has now
become a hot subdwarf, leaving the optical companion as a freshly contaminated
barium star inside an ionized planetary nebula.

\end{abstract}



\section{Introduction}

In this paper we report the discovery of a new planetary nebula (PN) in
Cassiopeia.  The PN appears nearly perfectly elliptical, and its central star
is a late-type star with enhanced abundances of carbon and \sprocess\
elements.  The system thus appears to represent the immediate aftermath of the
formation of a barium star.  In the following sections we present the
serendipitous discovery of the nebula, the classification of the nucleus as a
\BaII\ star, a study of the star's photometric properties and variability,
estimates of the distance to the system, and an evolutionary scenario for the
origin of this remarkable object.  We conclude with suggestions for follow-up
studies.

\section{The Planetary Nebula WeBo 1}

\subsection{Discovery}

In 1995, one of us (R.F.W.) discovered an unusual nebulous object while
verifying the coordinates of LS~I~+61$^\circ$303, a well-known stellar X-ray
source. The Digitized Sky Survey (DSS)\footnote{The DSS was produced at the
Space Telescope Science Institute under NASA grant NAG~W-2166. The images of
these surveys are based on photographic data obtained using the Oschin Schmidt
Telescope on Palomar Mountain.} was used to generate an image of
LS~I~+61$^\circ$303. During examination of the source and its surrounding
field, R.F.W. noticed a faint, elliptical nebula surrounding a
14th-mag star, lying $5'$ to the southwest of (and unrelated to)
LS~I~+61$^\circ$303. 

H.E.B. subsequently obtained narrow-band CCD images of the object, with the
0.9-m telescope at Kitt Peak National Observatory (KPNO) in autumn 1996,
and later with the KPNO Mayall 4-m telescope as described below.  Filters
isolating [\OIII] 5007~\AA\ and H$\alpha$+[\NII] 6584~\AA\ showed the object to
be a previously unrecognized\footnote{The SIMBAD database reveals, however,
that two other groups have independently noticed WeBo~1, but did not recognize
it as a PN\null. Hau et~al.\ (1995) included it in a list of candidate galaxies
lying at low galactic latitude. Mart\'\i\ et al.\ (1998) detected WeBo 1 with
the Very Large Array at 6 cm during their search for extended radio emission
associated with LS~I~+61$^\circ$303, and confirmed it optically using the DSS;
they suggested that it is a small \ion{H}{2} region.}~PN. 

The name WeBo~1 was proposed by Bond, Ciardullo, \& Webbink (1996), and will be
used here.  In the nomenclature of the {\it Strasbourg-ESO Catalogue of
Galactic Planetary Nebulae\/} (Acker et~al.\ 1992), which is based on galactic
coordinates, the object's designation would be PN~G135.6+01.0.  The central
star is listed in the USNO-A2.0 catalog at the J2000 coordinates given below in
Table~1. Unusually for a PN nucleus, the star appears to be quite red: the
USNO-A2.0 approximate photographic $B$ and $R$ magnitudes are 16.0 and 14.4,
respectively.

\subsection{Nebular Morphology}

Our best narrow-band images of WeBo~1 were obtained by H.E.B. on 1999 January
16 with the KPNO 4-m telescope and its Mosaic CCD camera.  The [\OIII] and
H$\alpha$+[\NII] images are shown in Figs.~1a and 1b, respectively. The frames
were taken through cirrus clouds, but under good seeing conditions: the FWHM of
stellar images in [\OIII] and in H$\alpha$+[\NII] was $1\farcs0$ and
$0\farcs8$, respectively.

In H$\alpha$+[\NII], WeBo~1 has a striking morphology, appearing as a nearly
perfect ellipse with major and minor axes of about $64''\times22''$. The shape
strongly suggests that the PN is a thin circular ring with a very low ratio of
height to radius, viewed at an inclination of $\sim$$69^\circ$.   Such a
morphology is nearly unique among PNe, matched only by the southern-hemisphere
PN SuWt~2, which has a nearly identical appearance  (Schuster \& West 1976; 
Bond, Exter, \& Pollacco 2001).  The H$\alpha$+[\NII] ring has a generally
clumpy appearance, a bright, sharp inner rim, and is brightest at the two ends
of its major axis (perhaps simply a path-length effect).  The interior of the
ring is almost hollow, although the brightness level is brighter than the
surroundings (which are overlain with diffuse H$\alpha$ emission at this low
galactic latitude).

In [\OIII] the PN appears more diffuse, but the ring is still apparent. Unlike
the H$\alpha$+[\NII] image, the interior of the ring is not hollow.  The images
suggest a gradient in ionization level, with the interior of the PN filled with
high-ionization material radiating in [\OIII], surrounded by cooler material
around the periphery of the ring emitting strongly in (presumably)
[\NII]\null.  Images through filters that separate H$\alpha$ and [\NII] would
shed further light on the morphology of WeBo~1.

\section{The Central Star}

\subsection{Classification as a Barium Star}

A CCD spectrogram of the central star of WeBo\,1 was obtained by D.L.P. at the
2.5-m Isaac Newton Telescope at La Palma, Canary Islands, on 1997 January 12.
The Intermediate Dispersion Spectrograph was used at a dispersion of
35.3~\AA~mm$^{-1}$. The spectral coverage was 980~\AA, at a resolution of
1.6~\AA, and the exposure was 1800\,s.

Fig.~2 shows the spectrum. To our surprise, the star is cool and chemically
peculiar. Molecular bands of C$_2$, CH, and CN are prominent, and the lines of
\SrII\ at 4077~\AA\ and \BaII\ at 4554~\AA\ are extremely strong.  Remarkably,
the core of the \CaII\ H line is filled by very strong emission. (\CaII\ K is,
unfortunately, just outside the spectral range that we covered.)

The star thus has all of the properties of a classical barium star (or ``\BaII\
star'').  \BaII\ stars were first identified as a spectroscopic class by
Bidelman \& Keenan (1951).  They show enhanced abundances of carbon and of
heavy elements produced by \sprocess\ neutron-capture reactions.  The modern
view of \BaII\ stars (see McClure 1984; Jorissen et~al.\ 1998; Bond \& Sion
2001) is that they are the binary companions of more massive stars that became
asymptotic-giant-branch (AGB) stars, dredged up C and \sprocess\ elements
from their interiors, and transferred them to the companions. The AGB stars
have now become optically invisible white dwarfs, leaving the contaminated
companions as the visible \BaII\ stars.


We classified the spectrum of WeBo~1 by displaying it as a ``photographic''
spectrogram, and comparing it visually with the atlas of Keenan \&
McNeil (1976).  To estimate the \BaII\ strength, which classically (Warner
1965) is on a scale from 1 (slightly enhanced over normal) to 5 (extremely
strong), we compared our spectrum with the sequence illustrated by L\"u et~al.\
(1983, their Fig.~2).  The resulting classification is K0~III:p~Ba5.  The
luminosity class is essentially indeterminate from our material (due to the
abnormal strength of all of the luminosity indicators at this resolution), and
the ``p'' refers to the presence of C$_2$ bands, which are generally seen only
in a subset of the \BaII\ stars, as well as to the \CaII\ emission.  


\subsection{{\it BVI\/} Photometry}

CCD {\it BVI\/} photometry of the nucleus was obtained by H.E.B. on six
occasions on four different photometric nights in 1996 and 1997 September. The
KPNO 0.9-m reflector was used with a Tektronix CCD, and the photometry was
reduced to the Johnson-Kron-Cousins {\it BVI\/} system through observations of
Landolt (1992) standard stars.  Two nearby stars were chosen as comparison
stars for a variability study (see below) and were also calibrated to the {\it
BVI\/} system.  Results are given in Table~1.  The errors, determined from the
internal scatter among the six different measurements, are $\pm$0.01~mag in $V$
and $B-V$, and $\pm$0.02~mag in $V-I$.

The central star of WeBo~1, at $B-V=1.72$ and $V-I=1.77$, is very red for its
spectral type; a normal K0~III star has $B-V\simeq0.98$ and $V-I\simeq1.00$
(Bessell 1979). The star thus appears to have an interstellar reddening of
$E(V-I)\simeq0.77$, corresponding to $E(B-V)\simeq0.57$ (using the formula of
Dean, Warren, \& Cousins 1978). The intrinsic color is $(B-V)_0\simeq1.15$,
suggesting that the star is somewhat redder in $B-V$ than a normal K0 giant
because of its strong carbon bands as well as the influence of the broad
Bond-Neff (1969) flux depression around 4000~\AA\ that is seen in barium stars.

Given that the optically visible star in WeBo~1 is very cool, we can speculate
that it is a member of a binary system with a much hotter, but optically
inconspicuous, companion that is responsible for the ionization of the PN.

\subsection{Photometric Variability}

In order to search for variability, H.E.B. obtained CCD images in {\it BVI\/}
on 40 occasions spread over 6 nights each in 1996 and 1997 September, 4 in 1998
March, and 2 in 1999 January.  The 0.9-m KPNO telescope was used, as described
above.  Since several of these nights were non-photometric, we calculated
differential magnitudes between WeBo~1 and the sum of the intensities of the
two comparison stars.

The central star is definitely variable, but at a low level.  We searched the
data for periodic variations, using the Lafler-Kinman (1965) algorithm.  The
smoothest light curves are obtained for a period of 4.686~days.  However, the
nearby alias periods at 4.347 and 5.074~days also give plausible light curves,
so we cannot make a definitive claim that 4.686~days is the correct period.  It
is even possible to fit the data with a sub-day alias near 0.82~days, but with
significantly larger photometric scatter.

In Fig.~3 we plot the differential magnitude and color curves for the
adopted ephemeris $T_{\rm min} = {\rm HJD} 2450346.6000 + 4.686 E$.   To estimate the uncertainties, we calculated the
standard deviations of the comparison star 1 minus comparison star 2
magnitudes, which are 0.011, 0.007, and 0.008~mag in $B$, $V$, and $I$
respectively.  The radii of the plotted points in Fig.~3 were then set to these
values (with the errors combined in quadrature for $B-V$ and $V-I$).
The WeBo~1 nucleus shows a nearly sinusoidal variation in all three filters,
with a peak-to-peak amplitude of about 0.03~mag.  There are no significant
changes in color.

We can identify four mechanisms that could produce this variation:
(1)~pulsation, (2)~heating (reflection) effects in a close binary,
(3)~ellipsoidal variations (at a period of $2 \times 4.7$ days), and
(4)~starspots on a rotating star.  Pulsation appears unlikely because of the
lack of any color variations.  For a planetary-nucleus cooling age comparable
to the estimated expansion age of the nebula (see below), we estimate its
luminosity should exceed $10^2 L_{\sun}$, in which case heating of the facing
hemisphere of the Ba star in a 4.7-day binary would produce a reflection effect
nearly two orders of magnitude larger than the observed photometric amplitude
(unless the binary is seen nearly pole-on), with colors decidedly bluer than
observed.  The reflection effect should also dominate ellipsoidal variability
at all inclinations, unless the barium companion is not the ionization source
for the PN\null.  We consider starspots the most likely explanation, in which
case 4.7~days is the stellar rotation period.  The strong \CaII\ emission
indicates that the star is very chromospherically active, supporting the
starspot interpretation and suggesting that the activity is dynamo-driven by
the relatively rapid rotation.

\section{Distance, Stellar Radius, and Rotation}

We can constrain the distance to WeBo~1 using three different methods.  The
first one uses the estimated interstellar reddening of $E(B-V) \simeq 0.57$ and
the three-dimensional extinction model of Arenou, Grenon, \& G\'{o}mez (1992).
The observed reddening is close to the asymptotic limit at large distances in
this region of the sky, but the Arenou et al.\ model allows us to set a
1$\sigma$ lower limit on the distance of 0.75~kpc, a result which is in
excellent accord with the extinction maps of Neckel \& Klare (1980).

An upper limit to the distance of WeBo~1 follows from equating the 4.7-day
photometric period with the rotational breakup period of the barium star. 
Assuming a mass of $1.5\,M_{\sun}$ for that star (Jorissen et al.\ 1998), we
find that its radius cannot exceed $13\,R_{\sun}$.  The observed magnitude and
reddening then imply that a K0~III star of this radius must be no farther than
2.4~kpc.

A third method uses a statistical distance scale for PNe.  There are several of
these in the literature, but we have chosen that of Cahn, Kaler, \&
Stanghellini (1992, hereafter CKS)\null.  Ciardullo et al.\ (1999) have shown
the CKS formalism to produce reasonably good agreement with PNe of known
distance, albeit with the factor-of-two scatter typical of all statistical PN
distance scales.  Adopting the 6~cm flux density of 4.4~mJy given by Mart\'\i\
et al.\ (1998), and an angular radius of $32''$, we use the CKS method to yield
a distance estimate of 2.2~kpc.

We adopt a nominal distance of $1.6 \pm 0.85$~kpc (the approximate mean and
range defined by the lower and upper limits).  At this distance, the absolute
visual magnitude of the central star is $M_V = +1.3^{-0.9}_{+1.6}$. Thus it has
roughly the luminosity of a normal red giant (in agreement with most of the
classical \BaII\ stars, e.g., Mennessier et al.\ 1997).  At this distance, the
major axis of the PN is $0.48 \pm 0.26$ pc, and the radius of the central star
is $9 \pm 5\ R_{\sun}$.  On the assumption of a nominal $20\kms$ expansion
velocity, the age of the PN is 12,000 $\pm$ 6,000~yr. If 4.7 days is the rotation
period of the \BaII\ star, then with a radius of $9\ R_{\sun}$ and an
inclination $i = 69^\circ$ (the apparent inclination of the PN), it would have
a projected rotational velocity $v \sin i \simeq 90\kms$. Since this is less
than the $\sim$$110\kms$ resolution of our spectra, we would not expect to see
an obvious spectroscopic signature, even though this would be about 2/3 of the
rotational breakup velocity.

\section{Discussion and Future Research}

WeBo~1 has several properties in common with the class of ``Abell~35''-type
planetary nuclei. This class was defined by Bond, Ciardullo, \& Meakes (1993),
and has been discussed by Bond (1994), Jasniewicz et~al.\ (1996),
Jeffries \& Stevens (1996), and Gatti et al.\ (1997).  In the three known
A~35-type nuclei (A~35, LoTr~1, and LoTr~5), a rapidly rotating late-type giant
or subgiant is seen optically, while a hot companion is detected at UV
wavelengths.  The cool components vary photometrically with periods of a few
days, corresponding to their rotation periods.  A definitive orbital period has
not been found for any of these three objects from radial-velocity studies,
suggesting that the orbital periods may be long.  This suspicion is confirmed
in the case of the field star HD 128220, which lacks a PN but is
otherwise similar in all respects to the A~35-type nuclei: its orbital period
is 872~days (Howarth \& Heber 1990).  These systems, then, have almost certainly
not undergone common-envelope interactions, which would have decreased their
orbital periods by substantial amounts.

As discussed by Jeffries \& Stevens (1996), there is a closely related class of
wide binaries containing hot white dwarfs and cool, rapidly rotating,
magnetically active {\it dwarfs}.  These authors propose a mechanism in which
an AGB star in a wide binary develops a dense stellar wind, part of which is
accreted by the companion star.  Their calculations suggest that significant
spin-up of the companion may occur, along with accretion of chemically enriched
material from the AGB star.  Although Jeffries \& Stevens considered their
suggestion somewhat speculative (in the absence of actual 3-D hydrodynamical
simulations of the accretion and spin-up), observational support has arisen in
the past several years.  This includes the finding of mild Ba enhancements in
the rapidly rotating dK component of 2RE~J0357+283 (Jeffries \& Smalley
1996), and in the nuclei of A~35 and LoTr~5 (Thevenin \& Jasniewicz 1997). 
WeBo~1, with its pronounced Ba and C overabundances, now provides further
support.

A scenario that emerges to explain the properties of WeBo~1 is thus as follows.
The progenitor system was a fairly wide binary whose components had nearly
equal masses (initial mass ratio $\sim$0.98).  The more massive star evolved to
the AGB stage, at which point the less massive component had also begun to
ascend the red-giant branch. The AGB star was constrained to rotate with the
orbital period, so that, as it developed a dense wind, the wind was ejected
preferentially in the orbital plane, leading to the ring-like nebular
morphology.  The wind was enriched in Ba and other \sprocess\ elements, and had
$\rm C/O>1$. A portion of the wind was accreted by the companion giant,
spinning it up to the observed 4.7-day rotation period, and contaminating its
photosphere with Ba, C, and other pollutants.  At present, the AGB star has
completely shed its envelope, exposing its hot core whose UV radiation ionizes
the ejected ring.


Several follow-up studies are clearly warranted. (1)~Ultraviolet spectra would
confirm the expected presence of a hot companion; a determination of its
surface gravity from its Ly$\alpha$ profile would provide a mass determination,
and thus an estimate of the progenitor's initial mass and  constraints on
evolutionary scenarios.  (2)~Radial velocities from ground-based spectra would
allow a search for the orbital period, or set a lower limit if, as we suspect,
the orbital period is long. (3)~High-dispersion spectra of the barium star
should be obtained, to determine its rotational velocity and atmospheric
elemental abundances.  In particular, it would be of great interest to search
for lines of technetium, which should be present if the star is really a very
recently created barium star.  (4)~An abundance analysis of the nebula would
also be of interest, since it may be possible to detect lines of heavy
\sprocess\ elements.



\acknowledgments

The discovery of WeBo~1 would not have been made without the DSS, and we
acknowledge the roles of the late B.M. Lasker and the many other persons
involved in creating this valuable resource.  KPNO is operated by AURA Inc.,
under contract with the National Science Foundation. We made use of the SIMBAD
database, operated at the CDS, Strasbourg, France. R.F.W. thanks B.D. Fields
for useful discussions, and acknowledges support from NSF grants AST 92-18074
and AST 96-18462. H.E.B. thanks R.~Gallino and D.~D.~Clayton for useful
suggestions, and W.~P. Bidelman for introducing him to \BaII\ stars more than
three decades ago.


\clearpage

\begin{figure}
\begin{center}
\includegraphics[height=3.9in]{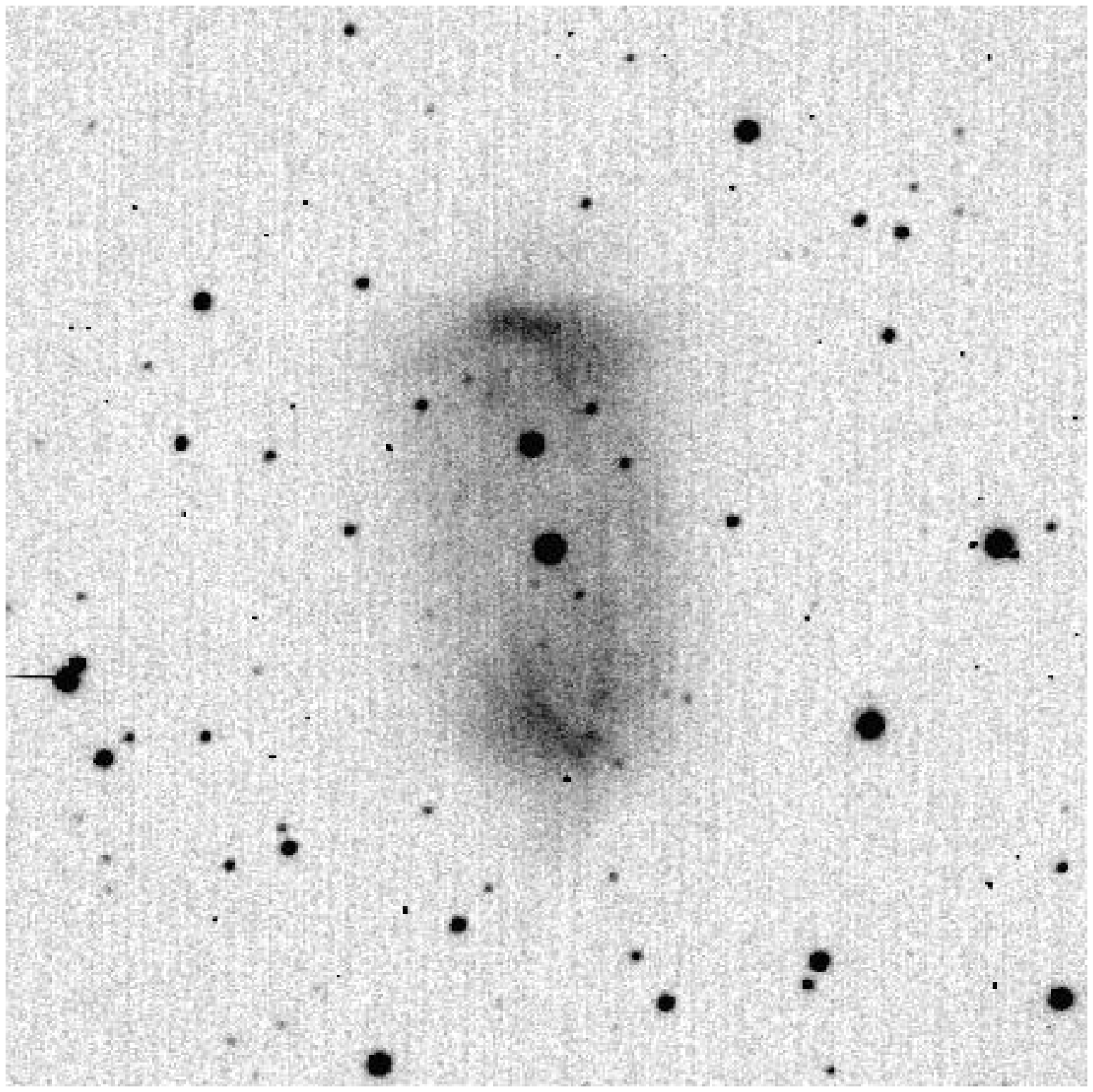}
\smallskip
\includegraphics[height=3.9in]{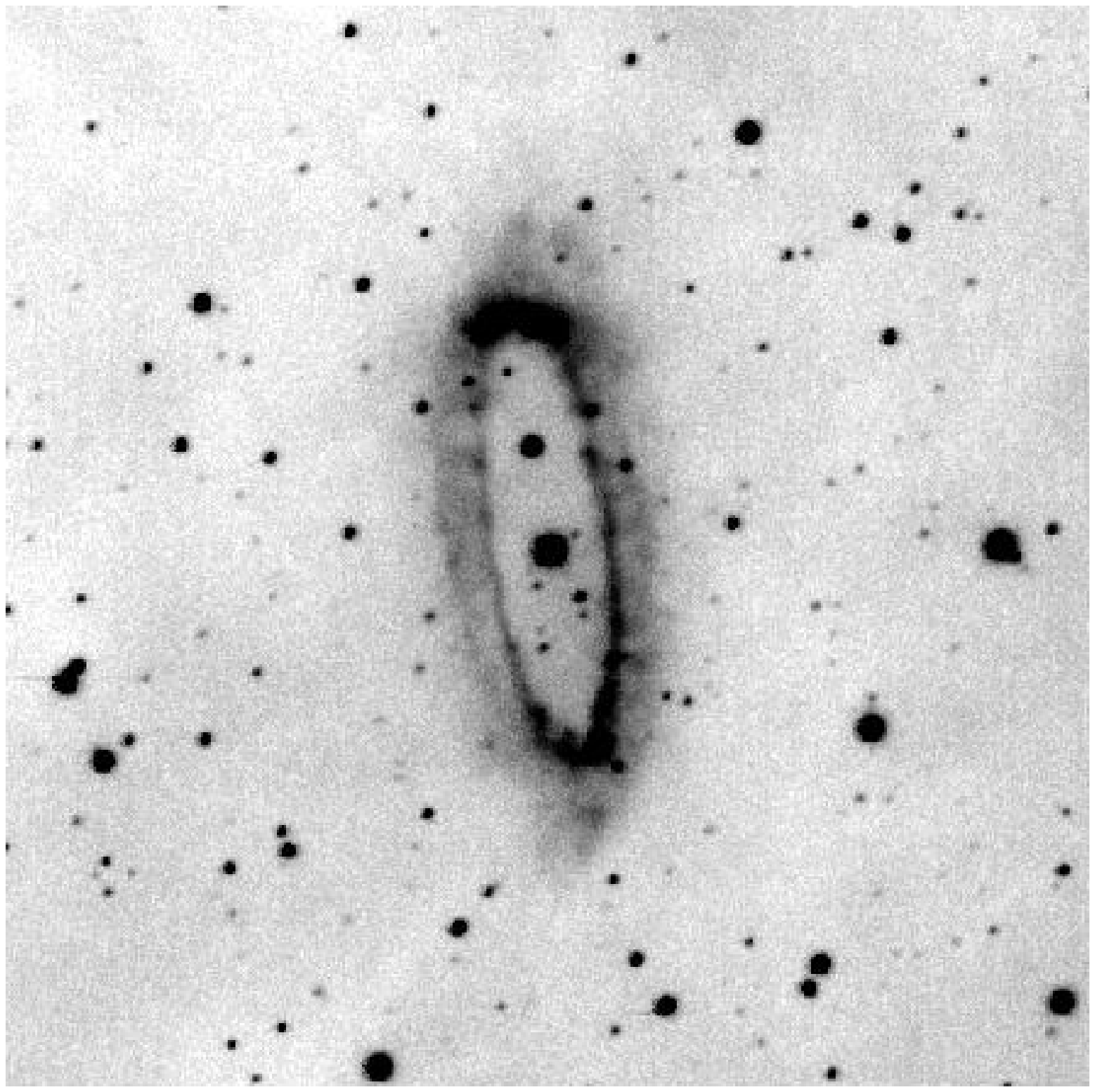}
\end{center}
\figcaption{KPNO 4-m CCD images of WeBo~1 in (a)~[\OIII] 5007~\AA\
and (b)~H$\alpha$+[\NII] 6584~\AA\null. North is at the top and east on the
left, and each frame is $150''\times150''$. The nucleus is the prominent
(14th-mag) red star lying at the center of the elliptical ring. Exposures were
$2\times300$\,s and $3\times300$\,s, respectively.}
\end{figure}
\clearpage

\begin{figure}
\begin{center}
\includegraphics[width=\hsize]{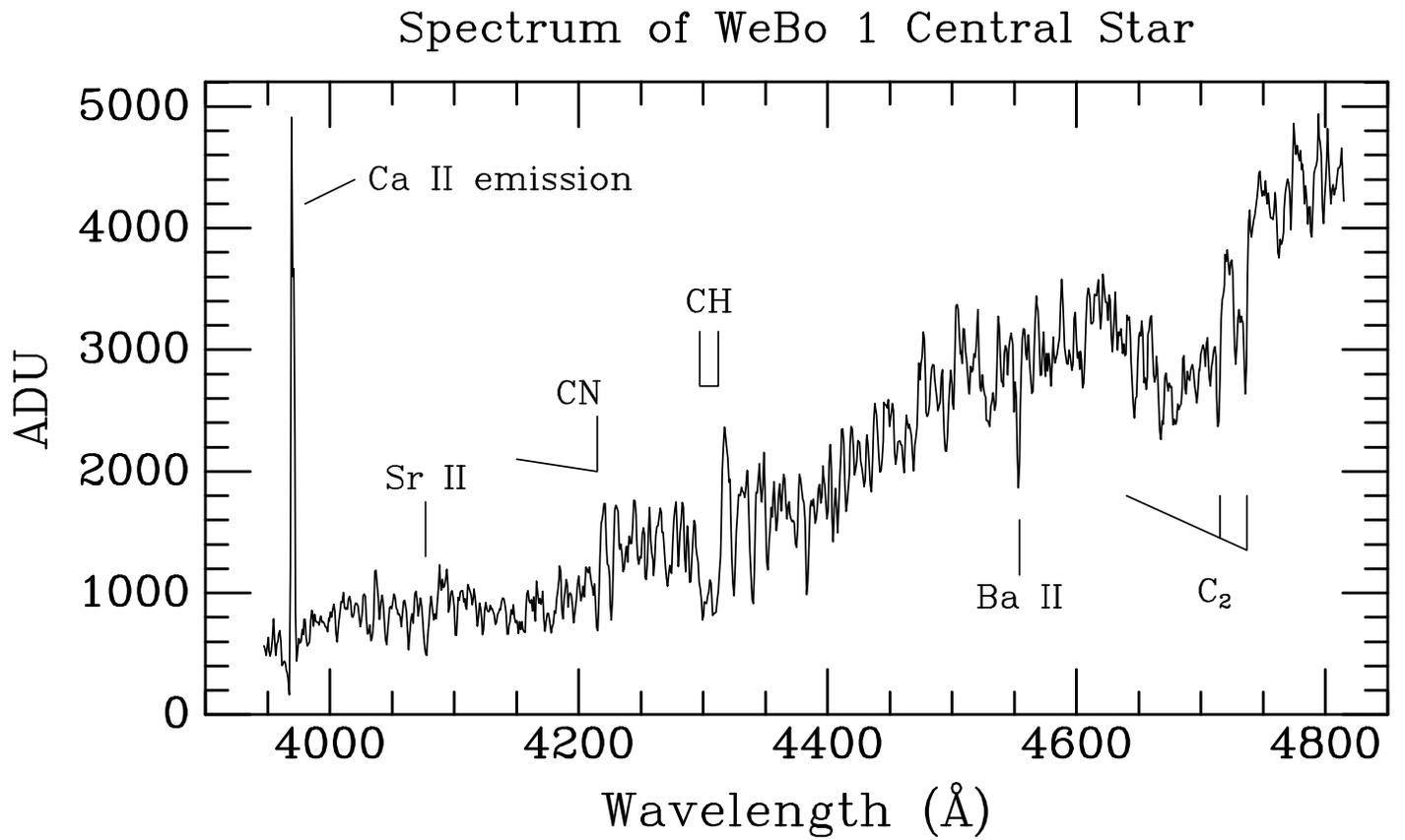}
\end{center}
\figcaption{INT spectrum of the central star of WeBo~1, with
several prominent spectral features marked. The star, which we classify 
K0~III:p~Ba5, is a classical \sprocess\ and carbon-enriched barium star, with
strong lines of \BaII\ 4554~\AA\ and \SrII\ 4077~\AA\, and strong bands of
C$_2$, CH, and CN\null. The \CaII\ H line is present as an extremely strong
emission line, indicating chromospheric activity.}
\end{figure}
\clearpage

\begin{figure}
\begin{center}
\includegraphics[width=5.5in]{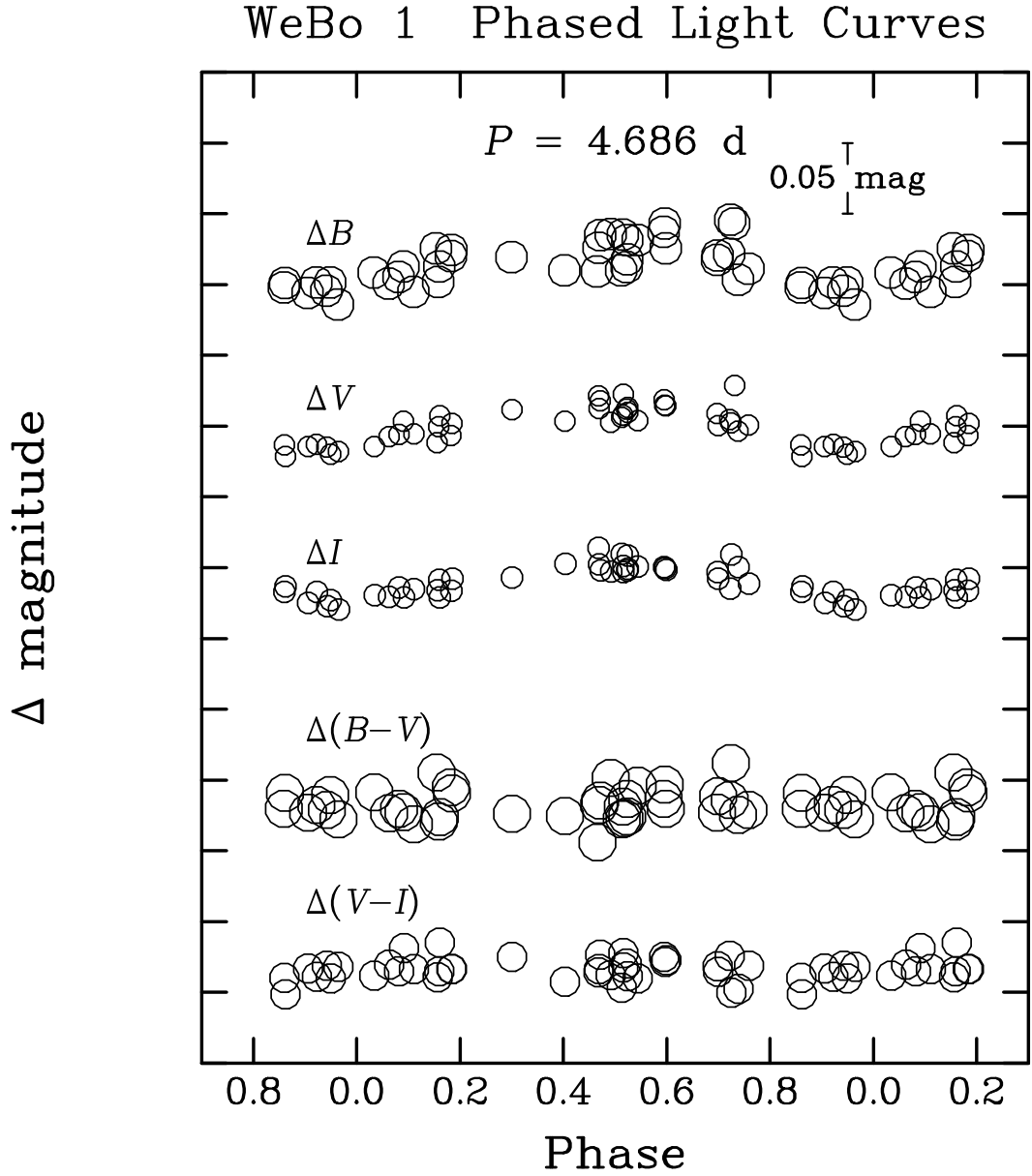}
\end{center}
\figcaption{Light curves of the nucleus of WeBo~1, phased with a period of
4.686~days. The differential $B$, $V$, and $I$ magnitudes, and differential
$B-V$ and $V-I$ colors, are plotted with arbitrary zero-point shifts.  Note
0.05~mag scale bar at the upper right. The radii of the plotted points have
been set to the estimated standard deviations of the magnitudes and colors.}
\end{figure}
\clearpage

\begin{deluxetable}{lcccccc}
\tablewidth{0pt}
\tablecaption{{\it BVI\/} Photometry of WeBo 1 and Comparison Stars}
\tablehead{
\colhead{Star} & \colhead{USNO-A2.0} & \colhead{$\alpha_{2000}$} &
\colhead{$\delta_{2000}$}  & \colhead{$V$}  & 
\colhead{$B-V$} & \colhead{$V-I$} }
\startdata
WeBo 1 Central Star & 1500-02588384 & 02:40:14.35 & +61:09:17.0 & 
	14.45 & 1.72 & 1.77 \\
Comparison 1        & 1500-02586253 & 02:40:05.78 & +61:09:17.9 &
	14.67 & 1.64 & 1.93 \\
Comparison 2        & 1500-02584109 & 02:39:56.80 & +61:10:03.8 &
	13.86 & 1.92 & 2.19 \\
\enddata
\end{deluxetable}

\end{document}